\documentclass[runningheads,a4paper]{llncs}

\usepackage[T1]{fontenc}

\usepackage{amssymb,amsmath,bbm}
\usepackage{graphicx}

\usepackage{url}
\urldef{\mailjr}\path|juste.raimbault@polytechnique.edu|

\newcommand{\keywords}[1]{\par\addvspace\baselineskip
\noindent\keywordname\enspace\ignorespaces#1}

\newcommand{\noun}[1]{\textsc{#1}}

\usepackage[normalem]{ulem} 
\usepackage{xcolor}
\newcommand\change{\bgroup\markoverwith
  {\textcolor{yellow}{\rule[-.5ex]{2pt}{2.5ex}}}\ULon}

\begin{document}

\mainmatter  

\title{\vspace*{-3cm}Reconciling complexities: for a stronger integration of approaches to complex socio-technical systems}

\titlerunning{Reconciling complexities}

%
%
\author{\noun{Juste Raimbault}$^{1,2}$}
\authorrunning{Reconciling complexities}

\institute{$^{1}$ UPS CNRS 3611 ISC-PIF, Paris, France\\
$^{2}$ UMR CNRS 8504 G{\'e}ographie-Cit{\'e}s, Paris, France\\
\mailjr
}


\maketitle

\begin{abstract}
Systems engineering has developed a mature knowledge on how to design, integrate and manage complex industrial systems, whereas disciplines studying complex systems in nature or society also propose numerous tools for their understanding. Socio-technical systems, that situate at their intersection, could benefit from a higher integration between these. This position paper advocates for such integrated approaches. A bibliometric study through citation networks first illustrates the respective isolation of some of these approaches. We then produce a proof-of-concept of how the transfer of concepts from biology can be useful for the design of complex systems, in the particular case of transportation networks, using a biological network growth model to produce various optimal networks in terms of cost and efficiency. We finally discuss possible disciplinary positioning of such hybrid approaches.
\end{abstract}

\section{Introduction}


Socio-technical systems can be understood as the appropriation and use of technical artefacts by social agents, and lie therefore at the intersection of engineered systems and complex social systems. Urban systems are a typical illustration of such systems and of the related issues to design and manage them \cite{portugali2012complexity}. According to \cite{sheard2009principles}, concepts and methods originating from complex systems science, such as self-organization, chaos or emergence, should much more frequently be used for engineering that they currently are, since engineers would be missing fundamental properties of the systems they design. This paper aims at confirming this claim through the use of case studies. We first precise what is meant by \emph{systems engineering} and \emph{complex systems science}.

\paragraph{Systems engineering}

The architecture of complex systems, in the sense of integrated systems of systems, is one objective of disciplines related to systems engineering. The International Council On Systems Engineering (INCOSE) is for example a major organization fostering the development of standardized practices and methodologies in that field, and gives regularly mature guidelines, such as for the architecture of systems of systems since a relatively long time \cite{maier1998architecting}. Model-based systems engineering (MBSE) \cite{estefan2007survey} is for example a methodology for designing systems, in which conceptual models of systems and data, together with their simulation, plays a central role. Several international standards have been introduced to structure practices \cite{schneider2013literature}. Current issues for the development of new methods include the coupling of existing methods with appropriated tools at large scales \cite{schafer2017challenges}.

\paragraph{Complex systems science}

There is not a single but several approaches to complexity in diverse fields of science, and we do not pretend to introduce a unified view of these. Indeed, \cite{chu2008criteria} recalls that several approaches to complexity do not currently converge to a single view. \cite{newman2011complex} gives an overview of concerned disciplines, that range from biology to physics and quantitative social science, and approaches, that include complex networks, agent-based modeling and simulation, non-linear dynamical systems. The common point of all the studied systems is to exhibit self-organization of a large number of elements, which translates into emerging properties at an upper level, in the sense of weak emergence which can be understood as the non-predictability of these properties, requiring simulation to understand the system \cite{bedau2002downward}. \cite{deffuant2015visions} actualize the metaphor of the Laplace Deamon, and develops three visions of complexity with progressive epistemological assumptions on the role of emergence, recalling that the computational complexity is already a barrier, but that higher orders of complexity, such as elementary constituting agents themselves complex, are often the rule in social systems.

\paragraph{Integrating complexities}


Similarly to \cite{sheard2009principles}, \cite{ottino2004engineering} advocates for a stronger consideration of emerging properties in the engineering of complex systems, and claims for example that engineers and social scientists have much to exchange. \cite{jennings2003agent} suggests that agent-based systems are an interesting alternative for the design of control systems, in particular thanks to their increased flexibility and robustness. These issues of integrating complexities is indeed not particular to system engineering, as \cite{farmer2009economy} show that in the case of economics, policy-related benefits would be obtained by a more frequent use of agent-based approaches. In the case of systems engineering, the architectured artefacts directly form components of socio-technical systems, and the integration is thus directly relevant. \cite{durantin2017disruptive} propose that the integration of systems engineering and design thinking could allow a smoother design process.

The aim of this paper is to confirm these views through the illustration by case studies. More particularly, our contribution is twofold: (i) we proceed to a bibliometric analysis of some branches of systems engineering and complex systems, and show that their connection exist but is negligible regarding their internal connections; (ii) we develop a modeling example in which the generation of a biological network is used to design transportation systems, and therein give a proof-of-concept of the possible transfer of concepts and methods.

The rest of the paper is organized as follows: we first develop a bibliometric study, in order to illustrate through the exploration of citation networks the effective separation of some branches of system engineerings and of complex systems science. We then develop a modeling case study to give a proof-of-study of how complex systems concepts, in this case from biology, can be used for the design of systems. We finally discuss the disciplinary positioning implied by higher levels of integration.

\section{A bibliometric insight}


Statements about disciplines, their positioning and their relations, must often be taken with caution, including ours, as they will depend on the perspective taken to enter the problem, on the information available, on possible higher contexts implying sociological issues \cite{latour1977rhetorique}. They furthermore involve issues of reflexivity if they are done by researchers in the field themselves, implying to find what Morin calls a ``meta-viewpoint'' to construct an integrated knowledge \cite{edgar1986methode}. However, a growing body of knowledge in bibliometrics \cite{waltman2010unified}, that can be understood as a \emph{quantitative epistemology} \cite{chavalarias2013phylomemetic}, allows to construct maps of knowledge, using scientific citations networks or other proxies of knowledge such as patents \cite{bergeaud2017classifying}. We take here this approach to gain a quantitative insight into the relation between the disciplines we consider.


We use the tool and method provided by \cite{raimbault2017exploration} to reconstruct backward citation networks from open data. More precisely, given an initial corpus, the scientific papers citing this corpus referenced in google scholar are obtained at a given depth (i.e. including the papers citing the papers citing, and recursively). We start from two nodes, namely \cite{estefan2007survey} as the origin node for system engineerings (considering thus the subdomain of model-based systems engineering approaches) and \cite{newman2011complex} for complex systems (considering an entry from physics). These two references constitute a relevant initial corpus for the following reasons: (i) both are surveys of the literature aiming at giving an overview, and have received a significant number of citations (552 for \cite{estefan2007survey} and 157 for \cite{newman2011complex}) which are comparable in orders of magnitude; (ii) they are not too specific and should represent a consequent part of the fields (in comparison, for complex systems, an alternative such as \cite{stone2000multiagent} would be targeted on multi-agents systems). These choices have naturally an influence on the final results, and we therefore do not claim to construct full maps of the disciplines but to quantitatively illustrate our issue. Programs, data and results of these analyses are available on the open git repository of the project at \url{https://github.com/JusteRaimbault/ComplexitiesIntegration}.

\begin{figure}
	\includegraphics[width=\linewidth,height=0.6\textheight]{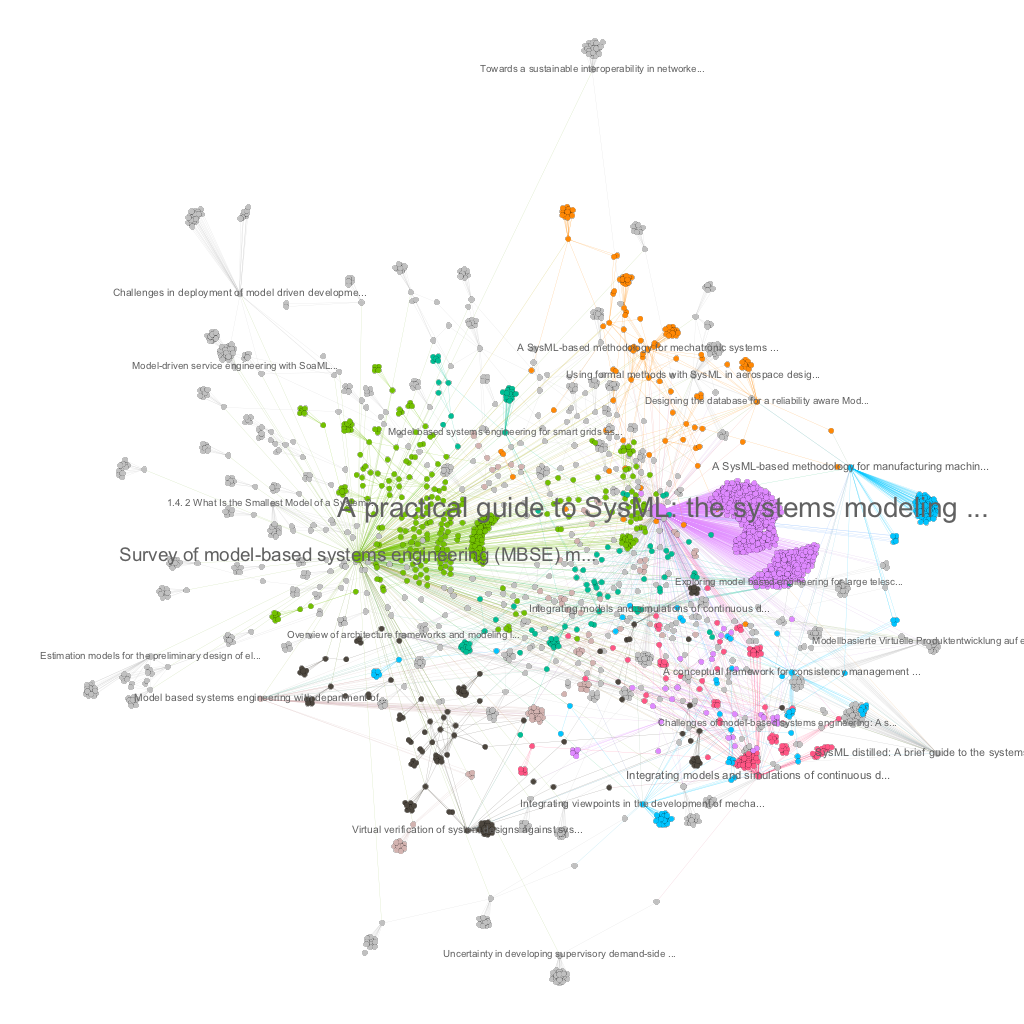}\\
	\includegraphics[width=\linewidth,height=0.6\textheight]{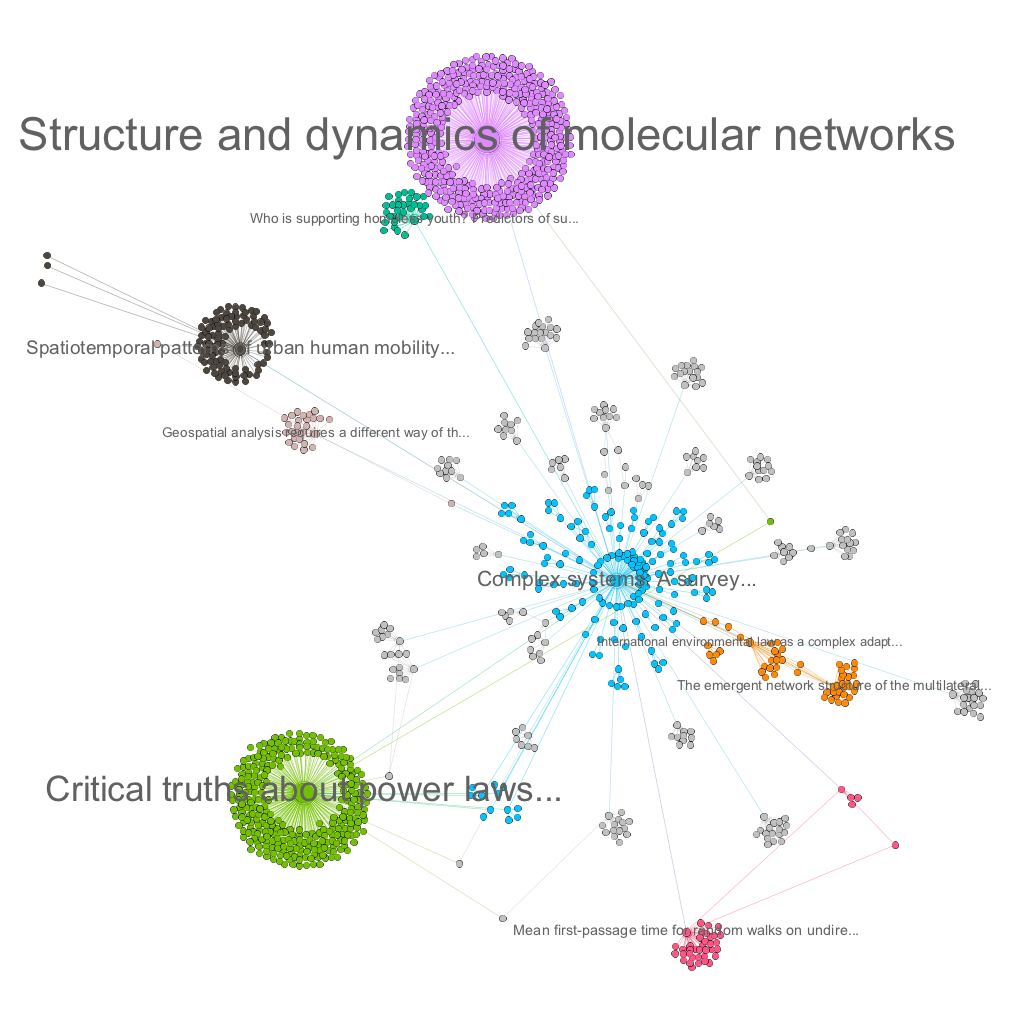}
	\caption{\textbf{Citation networks for each field.} \textit{(Top)} Citation network for the sample corresponding to system engineering approaches; \textit{(Bottom)} Citation network for the sample corresponding to complex systems science approaches.\label{fig:bothcitnw}}
\end{figure}

\begin{figure}[h!]
	\includegraphics[width=\linewidth]{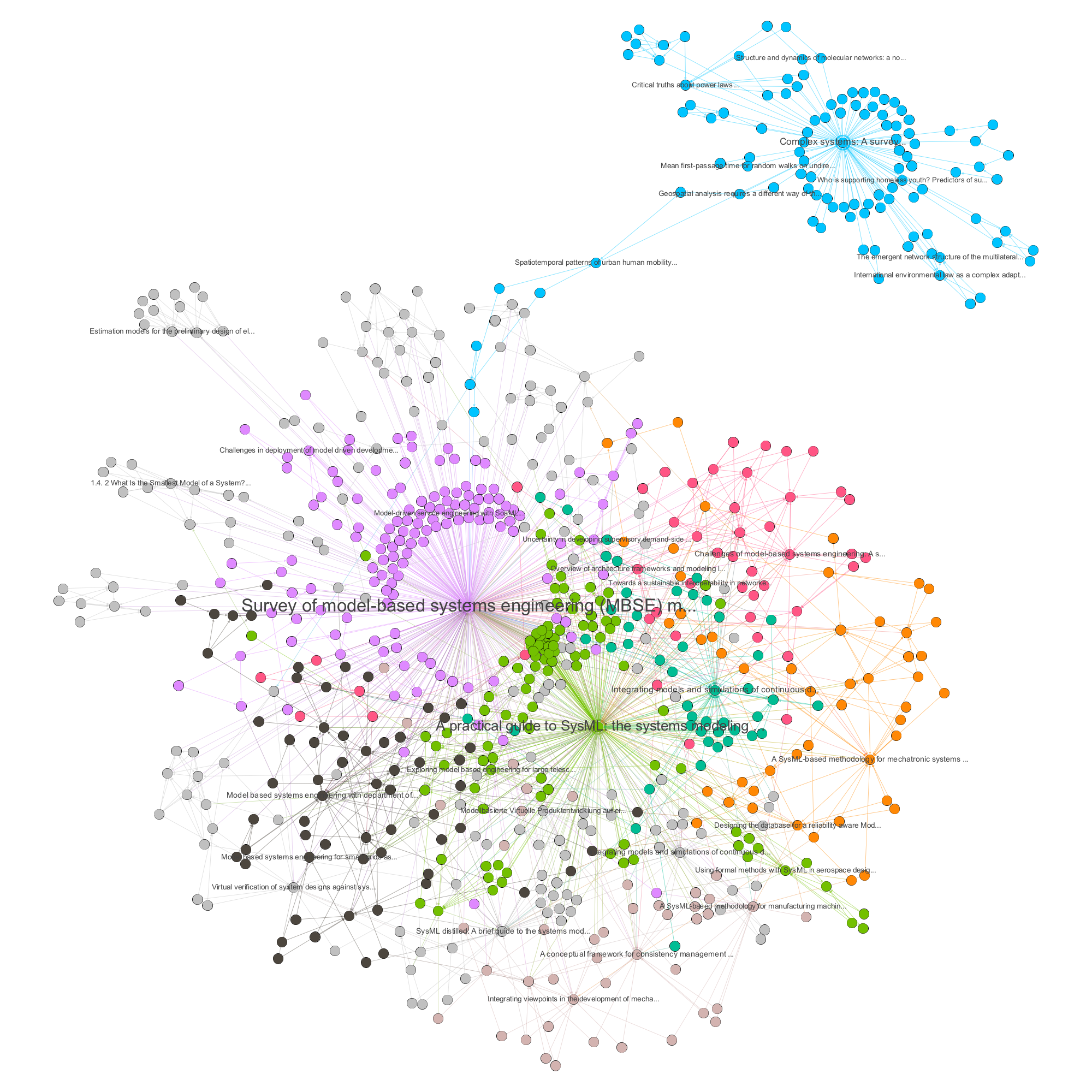}
	\caption{\textbf{Full citation network.} We show the core of the full citation network, corresponding to nodes with a degree larger than one and corresponding edges, to ease readability. The two components shown in Fig.~\ref{fig:bothcitnw} are connected by a small bridge only. Color gives the modularity class of nodes, and node size is proportional to degree.\label{fig:citnw}}
\end{figure}



The full citation network at depth 2 has 4019 nodes and 4183 edges, and a density of 0.003 which is standard for a citation network. The average in-degree of the core (obtained when removing nodes with a degree equal to one) is 2.13, what means that we indeed obtain branches of the disciplines considered. The network is highly modular, with a modularity of 0.59, corresponding to well identified subfields. We first visualize in Fig.~\ref{fig:bothcitnw} separately the two branches corresponding to the two initial nodes. The size of each is still comparable (3037 nodes for systems engineering and 1304 for complex systems). Clear communities, obtained with a standard community detection algorithm, can be distinguished within each branch: for example in systems engineering, MBSE has its own community; a second important being around the language SysML which is a programming language allowing the implementation of some MBSE approaches; other smaller communities corresponding to application-specific uses, such as manufacturing or aerospace. For the complex systems branch, we obtain various communities in biology, social science, economics, physics, geography. As expected, the latest is much more diverse in scope: we expect thus a higher chance of connexion between the two.

The core of the full network is shown in Fig.~\ref{fig:citnw}. Without tuning the resolution parameter of the community detection algorithm, we obtain a single community for complex systems and several for systems engineering, meaning that practices are comparatively much more clustered in the second. The fact that the two branches are connected (i.e. that the network has a single weakly connected component), was not necessarily expected, and confirms that bridges indeed exist. These are however tiny, as only two references lie in the intersection of the two branches. Surprisingly, the important paper in the intersection in terms of number of citations is \cite{hasan2013spatiotemporal}, which studies spatio-temporal patterns of human mobility: this confirms that socio-technical systems such as urban systems (of which mobility systems are a component) are indeed in the middle, and necessitate the integration of the two approaches.

Despite the limitation of this brief study, such as the scope fixed by the small initial corpus, we illustrate how systems engineering and complex systems do not communicate, and should have therefore their own distinct notions of \emph{complexity}. We propose in the following to show how bridges can be built in a concrete case.


%








\section{A proof-of-concept: biological network generation}

\subsection{Context}


How can the transfer of concepts from other disciplines can inform system design ? Contributions giving elements of answer to this question are indeed not new, and the entire field of \emph{Artificial Life} \cite{langton1989artificial} has indeed focused on these issues for long, since it aims at designing self-organized systems mimicking properties of living systems. Cellular automatons are a way to do it \cite{langton1986studying}, and properties such as autopoiesis (the ability of a system to maintain its network of processes by acting on its boundary conditions) and cognition have been suggested in Conway's game of life \cite{beer2004autopoiesis}. We consider here the generation of networks by biological entities, and more particularly a food transport network for a micro-organism called \emph{slime mould}, as described by \cite{tero2010rules} which gave a toy illustration of its application to a network design problem.

\subsection{Network generation model}

We detail here the model of type \emph{slime mould} used to evolve a biological network, introduced by~\cite{tero2007mathematical}. The broad idea of this model is that a flow, carrying food or people for example, circulates in a network of which the diameter of links evolve in time answering to the evolution of the flow. Self-reinforcement effects allow the convergence towards a steady state network.

A spatial network is composed by nodes characterized by their pressure $p_i$ and by links characterized by their length $L_{ij}$, their diameter $D_{ij}$, an impedance $Z_{ij}$ and the flow traversing them $\phi_{ij}$. The topology of the network is assumed fixed, but the diameters of links can evolve in time. We assume a grid topology with diagonals, in order to cover space with a fine resolution.

The flows in links are characterized by a relation analogous to Ohm's law which writes

\begin{equation}
\phi_{ij} = \frac{D_{ij}}{Z_{ij}\cdot L_{ij}} \left(p_i - p_j\right)
\end{equation}

Furthermore, the conservation of flows at each node (Kirchoff's law) imposes

\begin{equation}
\sum_i \phi_{ij} = 0
\end{equation}

for all $j$ except the source and the sink, that we assume at indices $j_+$ and $j_-$, such that $\sum_i \phi_{ij_+} = I_0$ and $\sum_i \phi_{ij_-} = -I_0$ with $I_0$ initial flow parameter. The combination of above constraints gives for all $j$

\begin{equation}
\sum_i \frac{D_{ij}}{Z_{ij}\cdot L_{ij}} (p_i - p_j) = \mathbbm{1}_{j=j_+} I_0 - \mathbbm{1}_{j=j_-} I_0 
\end{equation}

what simplifies into a matrix equation, by denoting $\mathbf{Z} = \left(\frac{\frac{D_{ij}}{Z_{ij}\cdot L_{ij}}}{\sum_i \frac{D_{ij}}{Z_{ij}\cdot L_{ij}}}\right)_{ij}$, and also $\vec{k} = \frac{\mathbbm{1}_{j=j_+} I_0 - \mathbbm{1}_{j=j_-} I_0 }{\sum_i \frac{D_{ij}}{Z_{ij}\cdot L_{ij}}}$ and $\vec{p} = p_i$. This yields

\begin{equation}
\left(Id - \mathbf{Z}\right) \vec{p} = \vec{k}
\end{equation}

The system admits a solution when $\left(Id - \mathbf{Z}\right)$ is invertible. The space of invertible matrices being dense in $\mathcal{M}_n(\mathbb{R})$, by multilinearity of the determinant, an infinitesimal perturbation of the position of nodes allows to invert the matrix if it is indeed singular. We obtain thus the pressures $p_i$ and as a consequence the flows $\phi_{ij}$.

The evolution of the diameter $D_{ij}$ between two equilibrium stages, i.e. two successive time steps, is then a function of the flow at equilibrium, through the equation

\begin{equation}
D_{ij} (t+1) - D_{ij} = \delta t \left[ \frac{\phi_{ij}(t)^\gamma}{1 + \phi_{ij}(t)^\gamma} - D_{ij}(t)\right]
\end{equation}

The default value for $\gamma$ is $1.8$, following the configuration used by~\cite{tero2010rules}. We furthermore take $\delta t = 0.05$ and $I_0 = 1$. We will vary only the parameter $\gamma$, what already gives a large flexibility on generated network configurations. The generation of a network can be achieved from an initial network, until reaching a convergence criteria, for example $\sum_{ij} \Delta D_{ij} (t) < \varepsilon$ with $\varepsilon$ fixed threshold parameter. We will use this model with a criteria of a fixed number of total iterations.

\subsection{Results}

The model is implemented in the NetLogo agent-based language, and an implementation as a generator of network for synthetic urban configurations is available at \url{https://github.com/JusteRaimbault/Governance/tree/master/Models/Lutecia}.

\subsubsection{Designing public transportation lines}

\begin{figure}[h!]
	\includegraphics[width=0.32\textwidth]{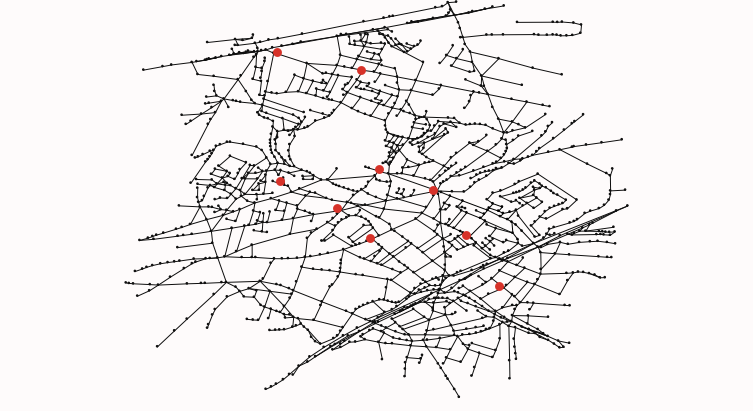}
	\includegraphics[width=0.32\textwidth]{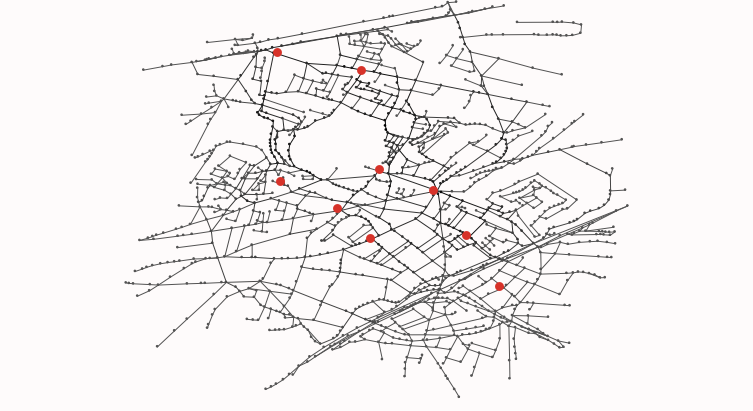}
	\includegraphics[width=0.32\textwidth]{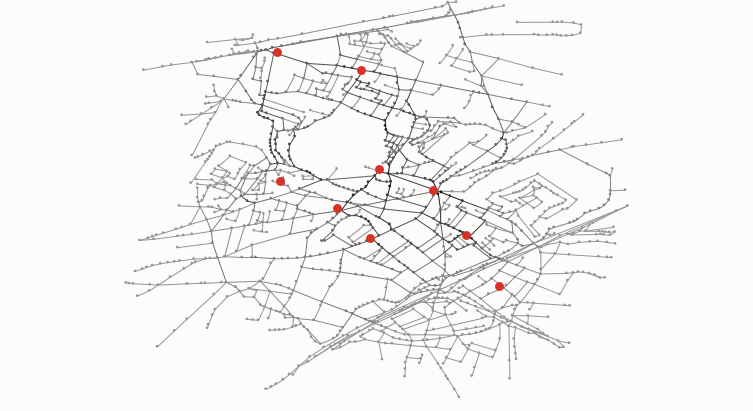}\\
	\includegraphics[width=0.32\textwidth]{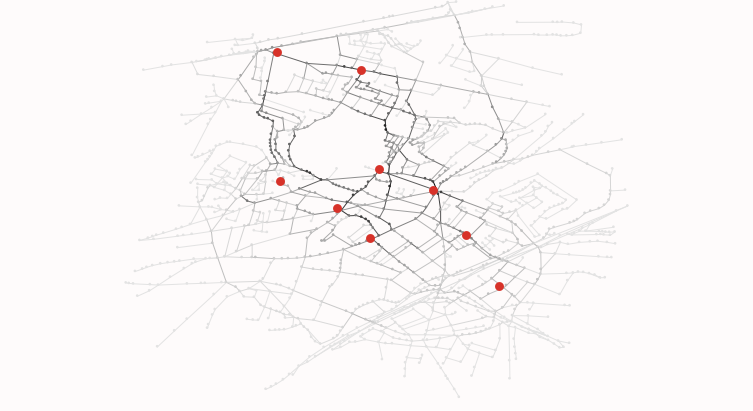}
	\includegraphics[width=0.32\textwidth]{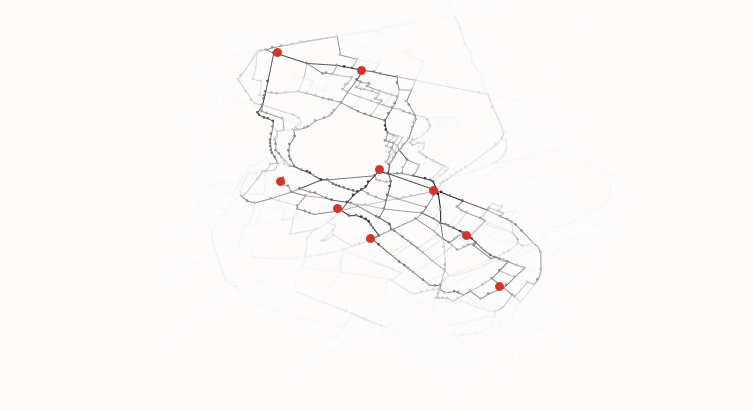}
	\includegraphics[width=0.32\textwidth]{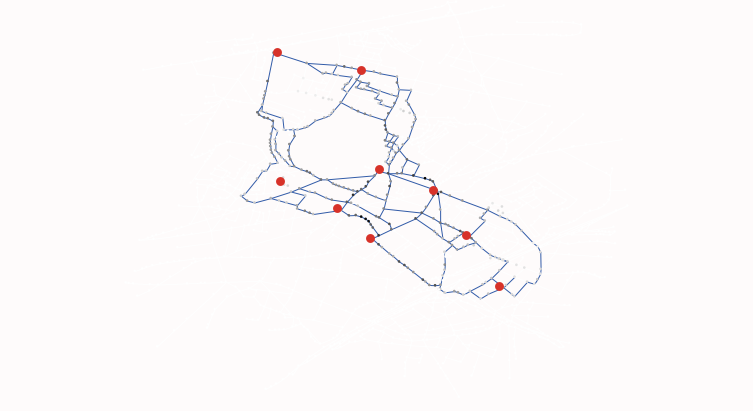}
	\caption{\textbf{Application of the slime mould model to design a robust public transportation network.} Given an initial network corresponding to the street network extracted from OpenStreetMap for the city of Romainville, France (top left), with stops in red which act randomly at each step as sources and sinks, the iteration of the model progressively converges towards a steady state with fixed diameters for links (from left to right and top to bottom).\label{fig:romainville}}
\end{figure}

A first illustration of an application of this model is the preselection of travelling routes for a bus transportation networks. Given a set of places to connect, route design is similar to a multi-objective travelling salesman problem \cite{angus2007crowding} since cumulated length of lines but also efficiency of the network must be optimized. Running the model on a street network with predefined stops will unveil preferred roads for the bus lines, as illustrated in Fig.~\ref{fig:romainville} for the city of Romainville, France.

\subsubsection{Generating optimal networks}

\begin{figure}[h!]
	\includegraphics[width=0.32\linewidth]{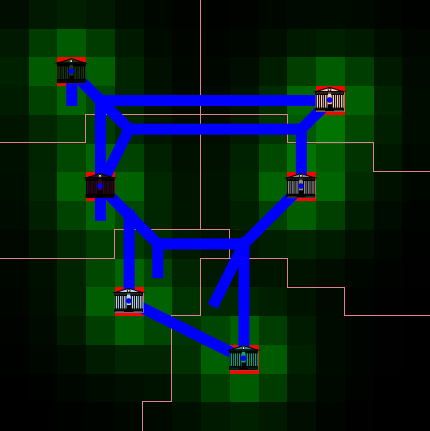}
	\includegraphics[width=0.32\linewidth]{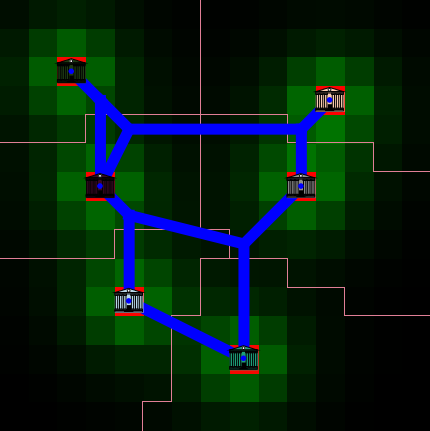}
	\includegraphics[width=0.32\linewidth]{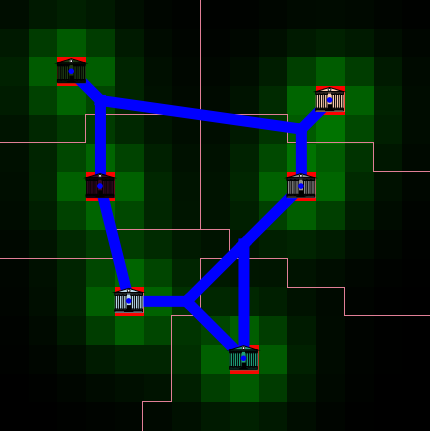}\\
	\includegraphics[width=0.32\linewidth]{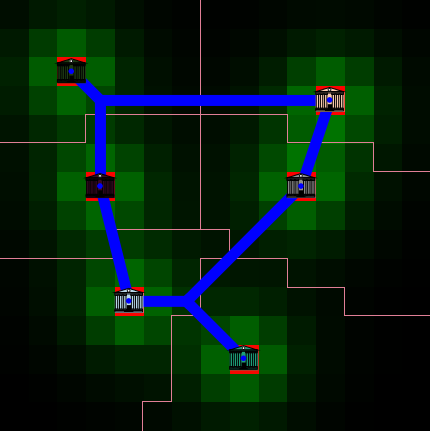}
	\includegraphics[width=0.32\linewidth]{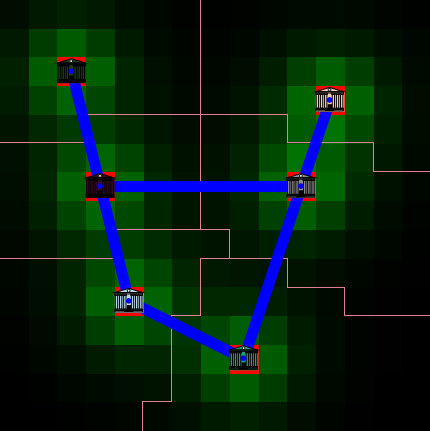}
	\includegraphics[width=0.32\linewidth]{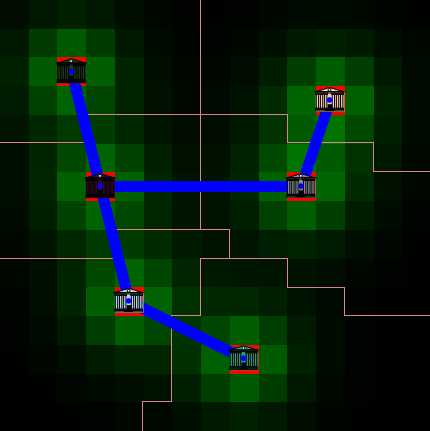}
	\caption{\textbf{Example of networks.} We show, for a given urban configuration (centers and population density in background color), the networks in blue generated with varying values of the parameter $\gamma \in \{1.1 ; 1.2 ; 1.25 ; 1.3 ; 1.5 ; 1.8 \}$ (from left to right and top to bottom).\label{fig:exnw}}
\end{figure}







\begin{figure}[h!]
	\includegraphics[width=1.2\linewidth]{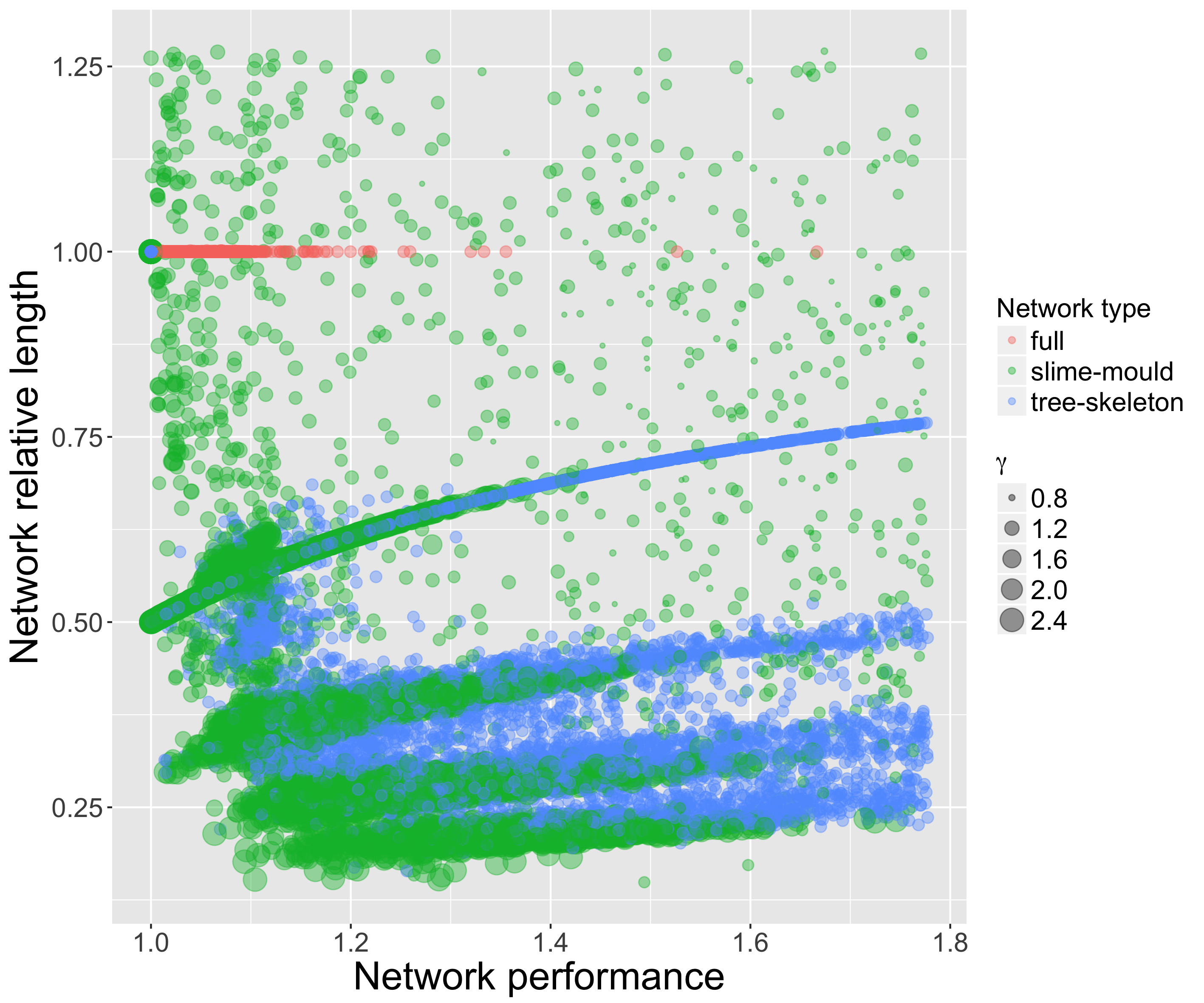}
	\caption{\textbf{Pareto evaluation of simulated networks.} We plot the values of the contradictory performance and length indicators, for all the simulated networks. Point color gives the model used to generate the network and point size the value of the reinforcement parameter $\gamma$ when the slime mould model is used.\label{fig:pareto}}
\end{figure}

The second application is to the generation of synthetic urban infrastructure network, for example for the automatic generation of scenarios for the test of Land-use transport interaction models \cite{wegener2004land}. Let assume that an urban configuration is given, in the sense of centers and an associated population distribution in space (which can be achieved for example using an exponential mixture model \cite{anas1998urban}). The issue of designing a reasonable infrastructure network associated to that structure, given constraints of optimality on cost, robustness and efficiency, is indeed a difficult problem. Given $N$ centers located in space, we generate a fine grid network which consists in the initial network for the slime mould model. At each time step, one center is chosen as the source with a flow $I_0$ and all others share the sinking flow with an equal proportion $-I_0 / (N - 1)$. The model is iterated a fixed number of steps $t_f = 1000$, after which nodes with diameter $D_{ij} < \varepsilon = 0.05$ are removed. The final network is then obtained by replacing all sequences of nodes with a degree of exactly two and the corresponding edge by a single link. We keep all parameters fixed except the reinforcement strength $\gamma$ which has a strong effect on emerging network structure.

We show in Fig.~\ref{fig:exnw} examples of generated network, for a fixed urban structure, obtained with different values of $\gamma$. Networks for lower values of $\gamma$ have more links, and are therefore more efficient in terms of connection speed between two arbitrary nodes and robustness to link deletion, but are more costly, whereas networks for higher values of $\gamma$ are cheaper but less robust. This shows the potentiality for a decision maker which has to make a compromise between these contradictory objectives. We proceed then to a more systematic exploration of the optimality of these network. We draw with a Latin Hypercube Sampling 100 random parameter values for the couple $(N,\gamma)$ such that $N \in \{ 2 ; \ldots ; 6 \}$ and $\gamma \in \left[ 0.5 ; 2.5\right]$, and we repeat a random uniform distribution of centers and the corresponding network generation 100 times. For a baseline comparison, two other network generation heuristics are used, namely a complete network between all the centers with direct links, and a tree network obtained by progressive connexion of closest weak components of the network. This gives a total of 30000 model runs. The significant corresponding computing time and the execution of this workflow is achieved using the OpenMOLE model exploration software, which allows a transparent distribution of model runs on a computation grid infrastructure \cite{reuillon2013openmole}. The indicators to be minimized within our experiment are the relative network length (sum of length of links divided by the length of a complete networks) and the relative performance (the average distance between nodes relative to the euclidian distance, see \cite{banos2012towards}). We show in Fig.~\ref{fig:pareto} the projected networks in the indicator space. The full network (red points) can have a good performance but is always very costly. The tree heuristic (blue points) is mostly dominated by the slime mould model (green points), since the majority of the corresponding point cloud lies at higher values of both indicators than a significant proportion of the slime mould point cloud. We observe that tuning the value of $\gamma$ strongly changes the location in the indicator space, allowing the production of a Pareto front for the two objectives. As expected and observed in the examples, low $\gamma$ values give costly but performant networks, whereas high $\gamma$ values give cheap and less performant networks.

We are through this experiment able to select Pareto optimal networks, answering a difficult engineering issue which is the design of a transportation network with multi-objective constraints. This example shows thus how the use of a model from a complex systems discipline can be useful to systems engineering.

\section{Discussion}


A more thorough exploration of the scientific landscape as we did in the second section, and a broader survey of bridging approaches as the one we illustrated, would give a more systematic demonstration of the idea we tried to defend in this paper. We however believe these case studies are additional clues of the lack of integration and of the potentialities offered by a stronger integration of approaches to complex socio-technical systems. The other direction, i.e. the use of systems engineering concepts in the study of complex systems, could be highly beneficial, as for example agent-based modeling has often the reputation of a lack of rigor \cite{rand2011agent}, and should be further investigated in a similar way.

A question directly raised by the perspective we give is to what extent such integrated approaches would belong to existing disciplines, or would they be intrinsically interdisciplinary, with all the complications and benefits it implies (see e.g. \cite{banos2013pour} describing the ``virtuous spiral of disciplinarity and interdisciplinarity''). This is a broad an open question, but we can at least give an example of a new discipline going beyond this distinction, which is \emph{Morphogenetic Engineering} \cite{doursat2012morphogenetic}. This rather novel discipline is a branch of Artificial Life, and focuses on designing and programming complex systems from the bottom-up. At the intersection between top-down architecting practices (typical of systems engineering) and bottom-up complex systems studies, it allows the elaboration of \emph{emergent architecture}. It remains disciplinary but is by nature integrated in its objects and methods, and is therefore an illustration of the kind of approaches we defend here.

The issue of which type of integration also remains open: the Complex Systems Roadmap \cite{2009arXiv0907.2221B} recalls that a \emph{vertical} integration must be done to create integrated disciplines tackling several levels of the systems considered, but that also an \emph{horizontal} integration in terms of questions transversal to disciplines being considered is necessary. We also suggest that an integration of \emph{Knowledge Domains} \cite{raimbault2017applied}, i.e. of the different components of knowledge such as theories, models, data, methods, would be highly beneficial to couple approaches that do not communicate yet, since these are the objects that can be borrowed for example. Finally, the issue of an operational implementation of such approaches, in particular for the systems engineering community which has strong constraints, is beyond the scope of this paper and should be investigated in future work.

\section{Conclusion}

We have in this paper given case studies to illustrate that systems engineering and complex systems approaches are highly disconnected, through a bibliometric study, and that the transfer of concept is possible, through the application of biological network generation to network design. This paves the way for similar more systematic studies enhancing reflexivity and the integration of disciplines.

\subsection*{Acknowledgments}

\footnotesize
Results obtained in this paper were computed on the vo.complex-system.eu virtual organization of the European Grid Infrastructure ( http://www.egi.eu ). We thank the European Grid Infrastructure and its supporting National Grid Initiatives (France-Grilles in particular) for providing the technical support and infrastructure.



\end{document}